\documentclass[%
 reprint,
 amsmath,amssymb,
 aps,
]{revtex4-2}

\usepackage{graphicx}% Include figure files
\usepackage{dcolumn}% Align table columns on decimal point
\usepackage{bm}% bold math
\begin{document}

\preprint{APS/123-QED}

\title{Astrometric detection of exoplanets in face-on orbits using vortex filters}% Force line breaks with \\
%\thanks{A footnote to the article title}%

\author{Niña Zambale Simon} \email{nfzambale@up.edu.ph}
\author{Miguel Revilla}
 %\altaffiliation[Also at ]{Physics Department, XYZ University.}%Lines break automatically or can be forced with \\
\author{Nathaniel Hermosa}%
\affiliation{%
 National Institute of Physics,University of the Philippines Diliman, Diliman, Quezon City, Philippines
}%

%\collaboration{MUSO Collaboration}%\noaffiliation

\date{\today}% It is always \today, today,
             %  but any date may be explicitly specified

\begin{abstract}
We propose a method to detect exoplanets based on their host star's intensity centroid after it passes thru a vortex filter. Based on our calculations with planets in face-on orbits, exoplanets with relative proximity to their host stars and with low mass ratios ($m_p/m_s$) can have discernible signals that can be amplified by the topological charge $\ell$ of the vortex filter.  For exoplanets that have higher mass ratios and are far from their host stars, a clear signal can also be obtained but these are not affected by the value of $\ell$ and other planet detection methods may be more suitable. We present a simple table-top optical experiment to support our calculations. Our proposed method adds to the arsenal of techniques for astrometric exoplanet detection.
%\begin{description}
%\item[Usage]
%Secondary publications and information retrieval purposes.
%\item[Structure]
%You may use the \texttt{description} environment to structure your abstract;
%use the optional argument of the \verb+\item+ command to give the category of each item. 
%\end{description}
\end{abstract}

%\keywords{Suggested keywords}%Use showkeys class option if keyword
                              %display desired
\maketitle

%\tableofcontents

\section{Introduction}

The Greeks in ancient times noticed that although stars have fairly predictable motion, there were a few of them, aptly called \emph{plan\=etes asteres} or wanderer stars, moving in the night sky peculiarly. Later on, these were shown to be completely different from stars in composition and movement. The observation of the path of these wanderers places astrometry, the precise measurement of the position and movement of celestial bodies, as the oldest branch of Astronomy \cite{evans1998history}. However, astrometry has been less successful in detecting exoplanets than other methods \cite{sozzetti2018space, launhardt2009exoplanet}. In fact, the first confirmed exoplanet detected using this technique—a planet orbiting a brown dwarf—was only discovered in 2013 \cite{sahlmann2013astrometric}. With the addition of the Global Astrometric Interferometer for Astrophysics (GAIA) spacecraft, the number of detected exoplanets had risen to five by early 2025~\cite{stefansson2025gaia}.

Astrometry is based on the motion of the host star about a common center of mass with its companion planet due to gravitational pulling. This motion depends on the mass of the planet $m_p$, the mass of the host star $m_s$ and the distance between the planets and the host star $d$ \cite{malbet2018astrometry,launhardt2009exoplanet,sozzetti2018space}. This motion is way too small ($\sim d\frac{m_p}{m_s}$) and its detection needs to be precise while correcting for other apparent motion of the host star - parallax and the proper motion of the star and its planetary system in the galaxy. Astronomers determine the planet's detection probability based on its possible astrometric signatures expressed as $\alpha=\frac{m_p}{m_s}\frac{a}{L}$, where $a$ is the semi-major axis, and $L$ is the distance of the host star from the observation point. The dependence on $L$ makes astrometry more suitable for nearby sources \cite{malbet2018astrometry, quirrenbach2010astrometric}. Moreover, a more accurate detection technique is needed for Earth-size planets in habitable zones, as astrometry has a bias on planets with large orbits \cite{malbet2018astrometry,launhardt2009exoplanet,sozzetti2018space, quirrenbach2010astrometric}.

There are of course more common exoplanet detection techniques such as the Transit Photometry (TP) \cite{batalha2014exploring, deeg2018transit, fischer2014exoplanet} and the Radial Velocity (RV) method \cite{hatzes2016radial,reiners2010detecting,knutson2014friends}. Almost 95\% of all the confirmed exoplanets have been detected with these two techniques. Unfortunately, these techniques leave a gap in detection when the planets have face-on orbits. No TP signal can be detected because the technique relies on the eclipse due to an orbiting planet and no radial Doppler shift in the light spectrum of the host star can be measured which the RV method hinges on. Thus, a considerable number of exoplanets may escape detection.  Direct imaging may solve this problem but it also carries issues such as the overwhelming brightness of the host star and the resolution limit dictated by the observing wavelength and the telescope size \cite{hecht2002optics,pueyo2018direct, launhardt2009exoplanet}. 

One solution to issues associated with direct imaging is the vortex nulling of the host star's light \cite{foo2005optical,tamburini2006overcoming, mawet2005annular, serabyn2010image, wagner2021imaging}. With this technique, Mawet et al demonstrated excellent contrast results of $\sim 2 \times 10^{-6}$ for $3\lambda/D$ angular separation where $\lambda$ is the wavelength and $D$ is the diameter of the telescope. However, Guyon et al calculated that coronagraphs are sensitive to the stellar angular size \cite{guyon2006theoretical}. Furthermore, they have quantified the performance limits imposed by Physics and have proposed designs to reach these limits. Unfortunately, Guyon et al have said that these design changes may be technologically difficult to implement \cite{guyon2006theoretical}. Small sample statistics can also play a role in coronagraphs' resolution limit especially when the planets are near the host star \cite{mawet2014fundamental}. Hence, it is crucial to develop a new technique for detecting planets in face-on orbits. 

In this article, we propose utilizing vortex filters, not to block the host star's intense light, but to transform it into a beam with an off-axis vortex. The Fraunhofer intensity distribution of the beam will change for a fixed vortex filter location, as the host star moves about its barycenter. Although the planet is not imaged, the changing centroid of the beam is related to the signature of a planet and is detected through a centroid detector such as a split detector or a quadrant detector automatically \cite{hermosa2011quadrant, narag2017response}. There is no need for a Lyot stop. This is akin to the signal detected for RV because of the gravitational interaction of the planet and its host star. But instead of a Doppler shift in the frequency, the centroid shift of the resulting Fraunhofer diffraction is observed.

This is not the first time that a beam with a displaced vortex filter was considered for astrometry \cite{thide2011applications,anzolin2009method}. Anzolin et al mentioned that the intensity difference between two opposite points of the beam with an off-axis vortex can be substantial at higher $\ell$ values when they analyzed the properties of the Fraunhofer diffraction pattern produced by a Gaussian light beam crossing a spiral phase plate \cite{anzolin2009method}. However, it was briefly mentioned only at the end of their conclusion and no other calculations were presented. Here, we provide analytical calculations and experimental demonstration that using a vortex filter enhances the intensity signal of a host star's centroid. Instead of two points in the beam as was done by Anzolin et al \cite{thide2011applications}, we integrate the intensity of the two halves of the diffraction pattern and subtract them-- a centroid detection \cite{hermosa2011quadrant, narag2017response}. By doing this, the signal is improved and there is no need to determine the intensity ratio between the two maximas \cite{thide2011applications}. Furthermore, we discuss how the vortex filter can be integrated in current experimentals setups, and provide the physical properties (size and mass) of exoplanets where the proposed technique proves effective. 

\section{Analytical calculations}

%and the more general case (c) when the barycenter is away by a distance $d_0$ and angle $\phi$ from the vortex and the detector center. In both configurations, the host star's waist is given by $w_0$.

\begin{figure}[ht]
	\centering
	\includegraphics[width=0.47\textwidth]{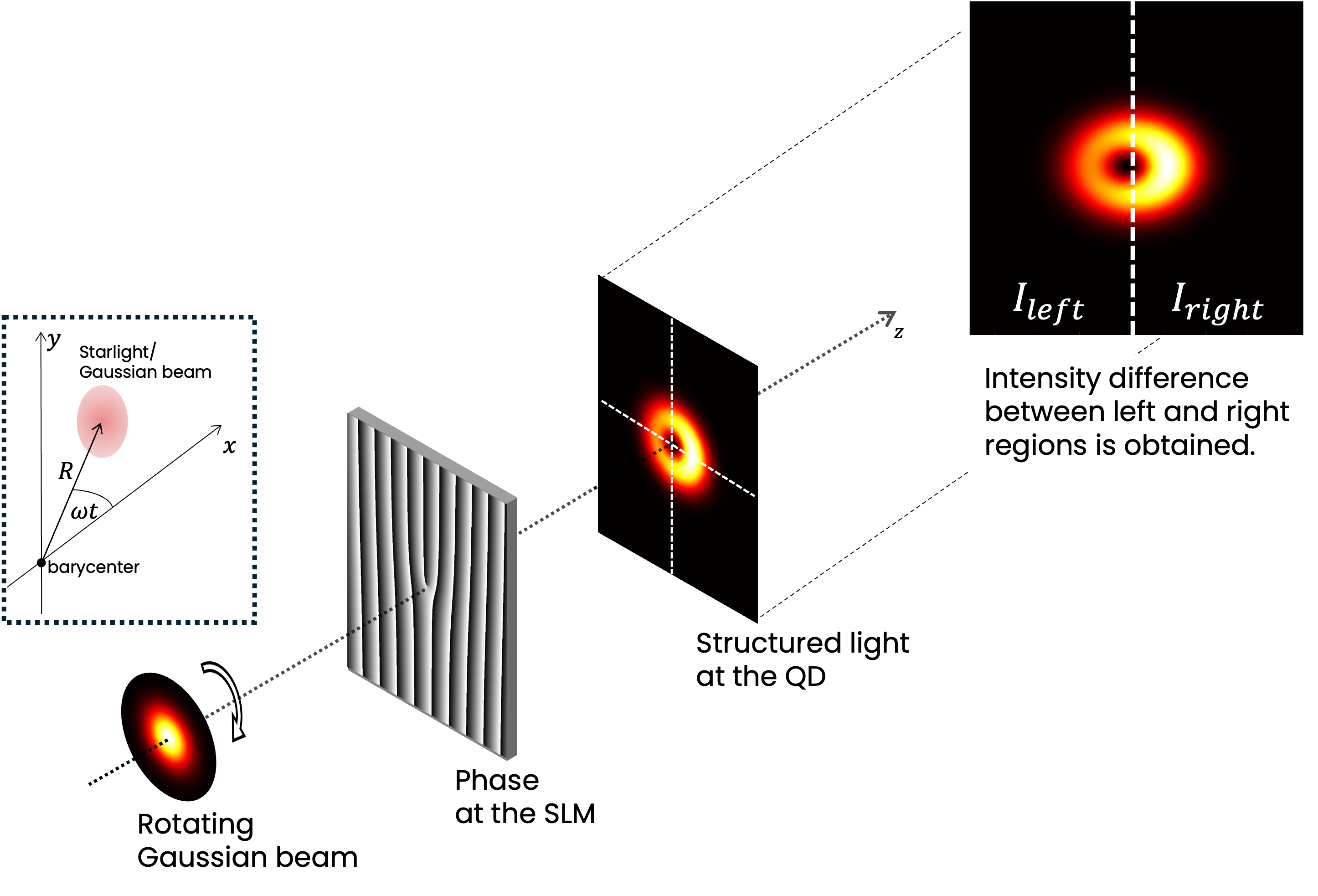}
	\caption{We consider a starlight, modeled as a Gaussian beam, passing thru a vortex filter and whose intensity is measured by a split detector. The vortex filter shares the same transverse coordinates as the split detector. We consider a case when the barycenter is aligned with the vortex and the detector center and the host star revolve with radius $R$. Inset shows the host star rotating around its barycenter due to the existence of an exoplanet.}
	\label{fig:system}
\end{figure}

We start our analytical calculations with the electric field of the light coming from the host star. With the existence of an exoplanet, the star-planet system then rotates about its barycenter which changes the direction of the starlight. We model the starlight from the host star as a Gaussian beam of waist $w_0$. In the presence of an exoplanet, the starlight gets displaced from its centroid as the host star now revolves around the star-planet's barycenter. We model this displacement as $R_x = R \cos \omega t$ and $R_y = R \sin \omega t$, where $R$ is the wobble radius or the distance of the host star from the barycenter, and $\omega$ is the star's angular frequency about the barycenter (\textit{see inset of Figure \ref{fig:system}}). We then let this starlight pass thru a vortex filter to gain an additional phase of $i \ell \Phi$ as expressed in Equation \ref{eqn:E_inc} where $\ell$ is the topological charge of the vortex filter, and $\Phi$ is the azimuthal phase. Figure \ref{fig:system} shows a visualization of the proposed technique of using a vortex filter. We consider a configuration wherein the barycenter is aligned with the vortex and the detector system. 

\begin{widetext}
	\begin{equation}
		E(\rho,\theta)=(\sqrt{2})^{|\ell|}\left(\frac{\rho}{w_0}\right)^{|\ell|}
		\exp\left\lbrace-\frac{\left[(\rho\cos\theta-R\cos\omega t)^2+(\rho\sin\theta-R\sin\omega t)^2\right]}{w_0^2}+i\ell\Phi\right\rbrace
		\label{eqn:E_inc}
	\end{equation}
\end{widetext}

To detect signal coming from the host star, we utilize a split detector which outputs the difference between its left and right quadrants. We are concerned with the signal difference in the left and right regions of the detector given by, 

\begin{align}
	\label{eqn:int_norm}
	\Delta I_\mathrm{norm} &= \dfrac{I_{right} - I_{left}}{I_\mathrm{tot}} \nonumber \\
	&= \dfrac{I(\rho, -\frac{\pi}{2}\leq\Theta\leq\frac{\pi}{2})-I(\rho,\frac{\pi}{2}<\Theta<\frac{3\pi}{2})}{I(\rho,\Theta)}
\end{align}

\noindent where $\rho$ is the radial coordinate with respect to the center of the vortex filter and the detector and $\Theta$ is the polar angle with respect to the center of the detector. To minimize variability in the intensity of the host star, the difference $\Delta I$ is normalized to the sum $I_{\mathrm{tot}}$. The intensity $I(\rho,\theta)=|E(\rho,\theta)|^2$ when $\Theta=\theta$, from Eqn. \ref{eqn:E_inc} is then given by 

\begin{equation}
\label{eqn: int1}
I (\rho,\theta)= 2^\ell\left(\frac{\rho}{w_0}\right)^{2\ell}\exp\left[-\frac{2(\rho^2+R^2)}{w_0^2}\right]\exp\left[\frac{4\rho R}{w_0^2}\cos(\theta-\omega t)\right]
\end{equation}

\begin{figure}[ht]
	\centering
	\includegraphics[width=0.5\textwidth]{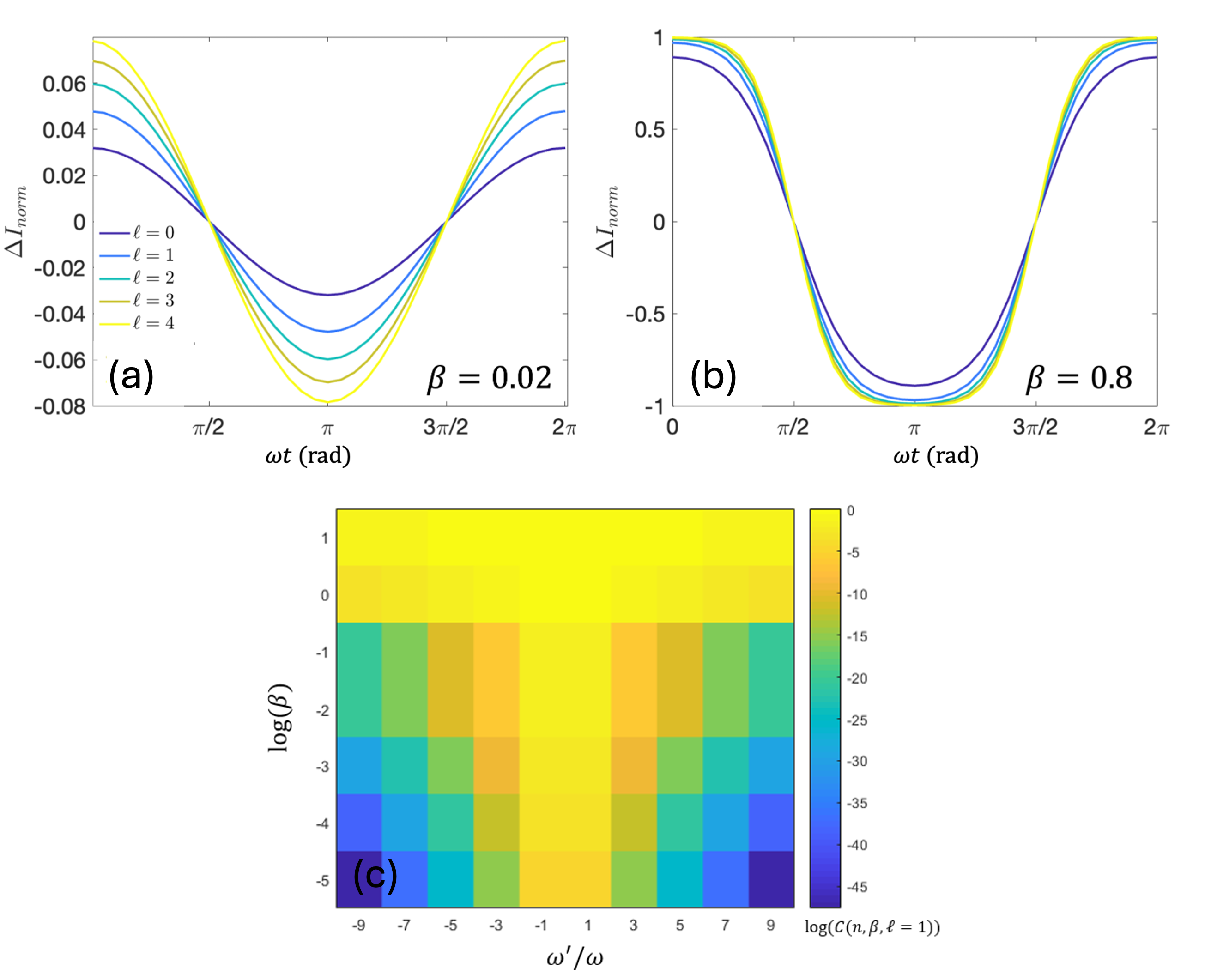}
	\caption{Plot of normalized intensity difference for different values of topological charge $\ell$ for (a) $\beta = 0.02$ and (b) $\beta=0.8$. Putting a vortex filter on the path of the starlight signal enhances the normalized intensity $\Delta I_{norm}$ for small $\beta$ values. This enhancement is directly proportional to the charge $\ell$ for small $\beta$. This is not the case for large $\beta$ where there is minimal enhancement. (c) The relative amplitude of the factor $C(n, \beta, \ell =1)$ following a Fourier transform of the $\Delta I_{norm}$. For small $\beta$s the first term is enough while large $\beta$s need more terms to approximate the curves. However regardless of the frequency, the $\beta$ values determine the n values.}
	\label{fig:deltaI}
\end{figure}

\noindent We used Eqn. \ref{eqn:E_inc} as in \cite{asymmetricgaussian} to ease calculations, compared to the Kummer equation used when a beam impinges a vortex filter \cite{anzolin2009method,bekshaev2008spatial}. While it may be easier to do numerical calculations using both and/or the Kummer equations, an analytical expression automatically relates the variables $R$, $\theta$, $w_o$, and the signal detected $\Delta I_{norm}$ by the split detector. Perfoming a change of variables $\rho' = 4\rho R/w_0^2$ and $\theta'=\theta-\omega t$ \cite{AbraSteg} in Eqn. \ref{eqn: int1} yields,  

\begin{widetext}
\begin{equation}
\label{eq:DeltaII}
    \Delta I_\mathrm{norm}=\sum_{\textnormal{n odd}}\frac{(-1)^{(n-1)/2}2^{(n+2)/2}\beta^n \Gamma\left(\frac{2\ell+2+n}{2}\right){}_1F_1\left(\frac{2\ell+2+n}{2};n+1;2\beta^2\right)}{n\pi\Gamma(n+1)\Gamma(\ell+1){}_1F_1\left(\ell+1;1;2\beta^2\right)}\times \cos(n\omega t).
\end{equation}
\end{widetext}

\noindent where $n$ is an integer greater than 0, ${}_1F_1(a;b;z)$ is the hypergeometric function, $\Gamma(m)$ is the Gamma function, and $\beta = R/w_0$ is the ratio of the wobble radius $R$ to the size of the star $w_0$ on the detector. This \textit{wobbling ratio} $\beta$ is related to the interaction of the exoplanet and the host star. Equation \ref{eq:DeltaII} represents the first key result of this study. From this equation, we see that the cosine term in the normalized signal manifests the periodicity of the star light about its barycenter. The equation also explicitly reveals the dependence of the $\Delta I_{norm}$ to the topological charge $\ell$ and the wobbling ratio $\beta$. By plotting Equation \ref{eq:DeltaII}, we observe that the presence of the topological charge amplifies the normalized signal compared to that of a Gaussian beam ($\ell = 0$) as presented in Figure \ref{fig:deltaI}(a). However, this enhancement is pronounced only for small values of $\beta$. For larger $\beta$, increasing $\ell$ values does not enhance the signal (Figure \ref{fig:deltaI}(b)).

Owing to the $\beta$ functionality of Equation \ref{eq:DeltaII}, we expect that the summation $n$ will be terminated at higher values for large $\beta$. We show this by taking the Fourier transform, $\widetilde{\Delta I_\mathrm{norm}}(\omega') = \sum_{\textnormal{n odd}} C(\beta,n,\ell) [\delta(\omega'+n\omega)+\delta(\omega'-n\omega)]$ where the cosine terms will give Dirac deltas centered at $\pm n\omega$. The factor $C(\beta,n,\ell)$ determines the $n$ values with which we can terminate the summation without sacrificing precision. Figure \ref{fig:deltaI}(c) shows that we can end at the first term for small $\beta$ ($\beta<< 1$). It will fit an almost perfect cosine function. The 1st term dominates while the succeeding terms are negligible. However, for larger $\beta$, higher-order terms are necessary.  This can be seen from  the same figure (Figure \ref{fig:deltaI}(c)) where $\beta\approx 10^{0}-10^1$  have significant second and third terms.  Fortunately, the frequencies are coupled to the $\beta$ values by the $n$ values alone and they are just odd integers of the primary frequency, $\omega$. We will discuss later the implications on exoplanet detection.

\section{Experiment}
%discuss how the starlight can be coherent?

\begin{figure}
	\centering
	\includegraphics[width=0.5\textwidth]{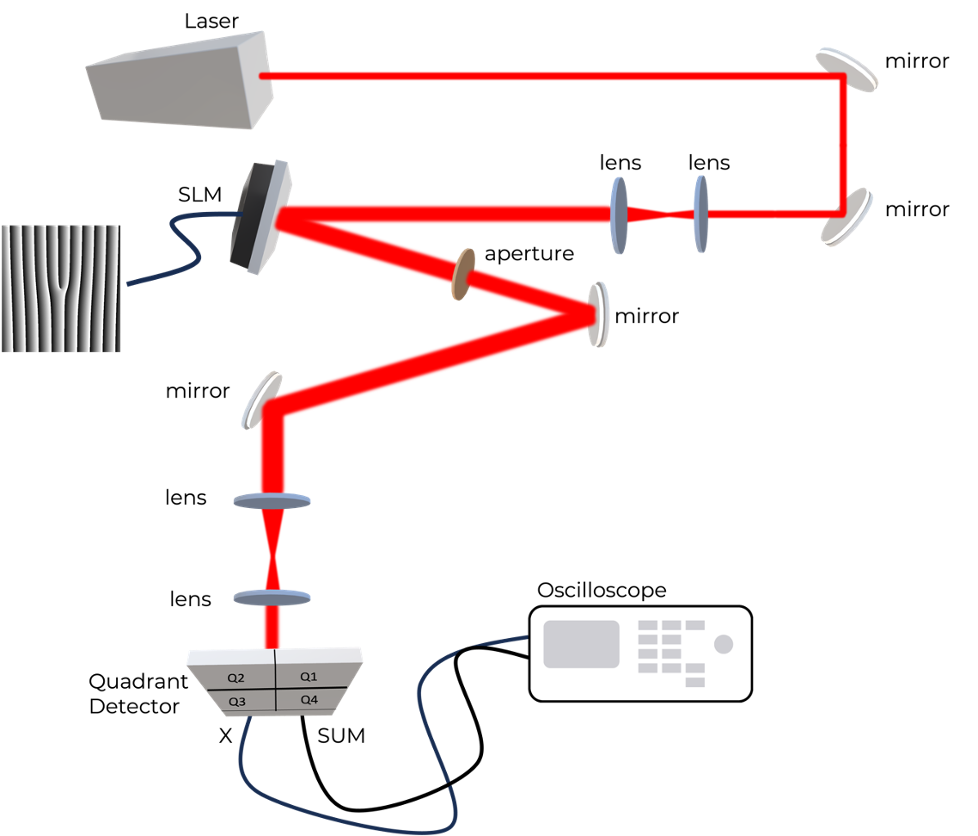}
	\caption{\textit{Experimental setup.} A 632-nm laser simulates starlight, which is directed onto the surface of a spatial light modulator(SLM) where a vortex filter is uploaded. As an exoplanet orbits, the starlight rotates around the barycenter of the system. To mimic this motion, the incident beam is made to rotate about the axis of the vortex filter. The resulting beam is imaged using a 4$f$ optical system and received by a quadrant detector(QD) attached to an oscilloscope. } 
    \label{fig:setup}
\end{figure} 

\begin{figure}
	\centering
	\includegraphics[width=0.38\textwidth]{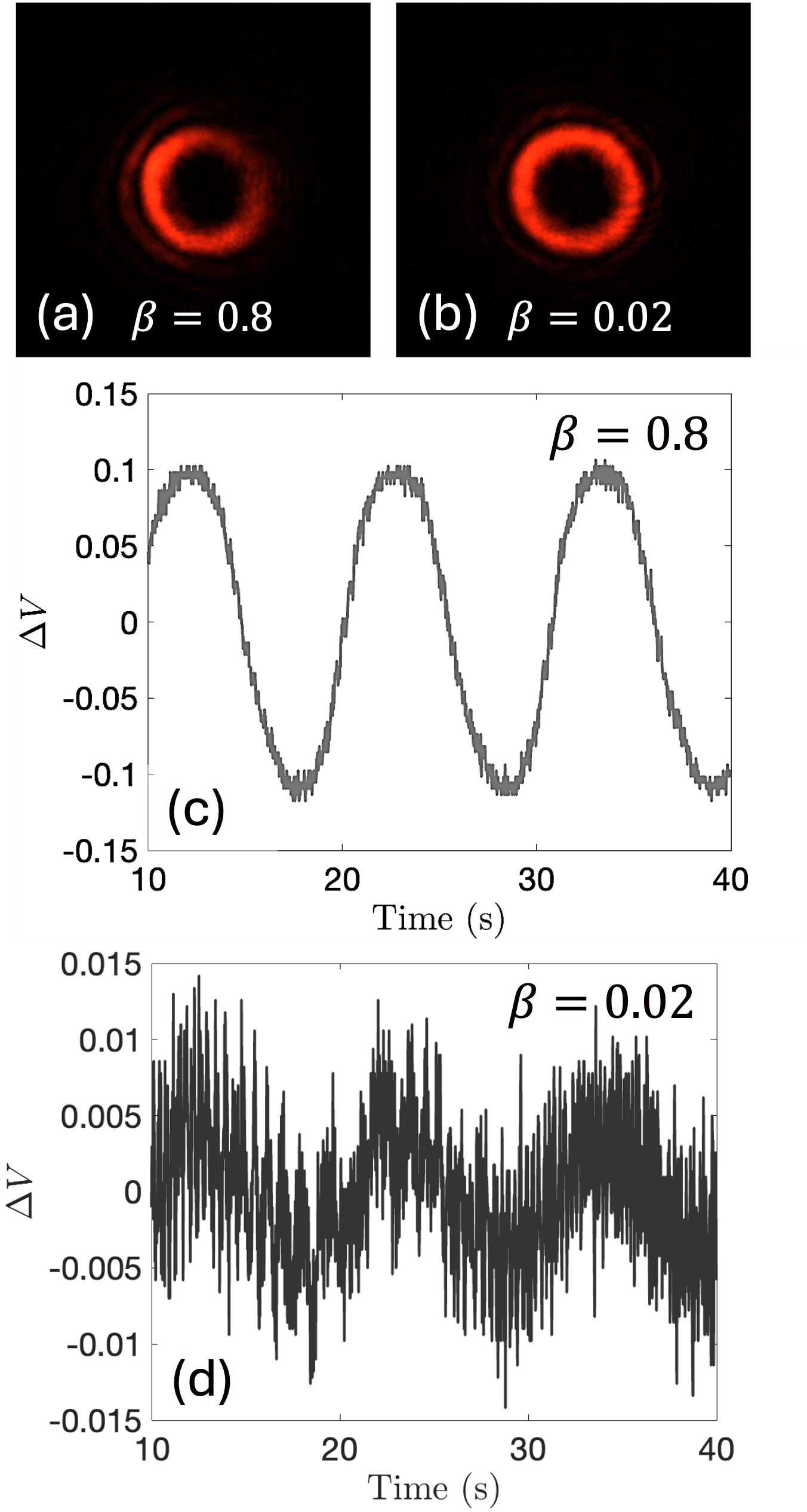}
	\caption{\textit{Sample experimental results.} The generated asymmetric vortex beam ($\ell = 5$) at the quadrant detector (a) with $\beta = 0.8 $ and (b) with $\beta = 0.02$ as captured by a CCD camera. Having a nonzero $\beta$ value results to an uneven intensity distribution of the beam. However for small $\beta$, this asymmetry is hard to observe thru a camera. The acquired raw signal from the oscilloscope for (c) $\beta = 0.8$ and (d) $\beta= 0.02$ showing the sinusoidal change in the signal from the quadrant detector. Because of the very small distance between the barycenter and the axis of the vortex filter, the signal is more noisy for $\beta=0.02$}
    \label{fig:results}
\end{figure}

In order to observe the normalized intensity difference $\Delta I_{norm}$, we design an experiment to measure the response of the split detector to different topological charge $\ell$ values. Figure \ref{fig:setup} shows the schematic diagram of the table-top experiment to determine if the vortex filters can be used to enhance $\Delta I_{norm}$. In our setup, the starlight is assumed to be a Gaussian beam coming from a 632-nm laser source. We then insert a vortex filter onto its path with topological charge $\ell$ using a spatial light modulator (SLM). As the star-planet system rotates about its barycenter, the direction of the starlight changes. We simulate this movement by letting the incident Gaussian beam move around the axis of the vortex phase and the quadrant detector. We vary the displacement of the Gaussian beam by $R_x = R \cos \omega t$ and $R_y = R \sin \omega t$, similar to the definition in the previous section. Experimentally, we set $\omega t$ equal to $0$ to $2\pi$ to simulate the periodic movement of the exoplanet, and the value of $R$ is dependent on $\beta = R/w_0$ with $w_0 = 1.6~\mathrm{mm}$. Figure \ref{fig:results}(c) and (d) show sample raw data of voltage difference $\Delta V$ recorded by the quadrant detector for $\beta=0.8$ and $\beta=0.02$, respectively. The obtained data in \ref{fig:results}(d) is noisy since the induced displacement to the incident beam is very small ($\beta = 0.02$), corresponding to just $2\%$ of the beam waist. Despite the noise, the sinusoidal trend of $\Delta V$ versus $\omega t$ is still evident. The quadrant detector is sensitive enough to detect minute displacements of the incident starlight. As for $\beta=0.8$, the signal noise is lesser owing to the larger displacement. Figure \ref{fig:results}(c) and (d) resembles figure \ref{fig:deltaI} (a) and (b).

%calibration of the QD. 
Due to its high sensitivity, the quadrant detector (QD) is used to perform position sensitive detection. The QD consists of four photodiodes with very small separation between each region. The calibration response of the QD has been measured for symmetric beams such as a Gaussian beam, Laguerre Gaussian beam, hard-ringed beams and Hermite Gauss beams~\cite{hermosa2011quadrant,narag2017response}. Particularly for vortex beams, the calibration constant $K$ is a function of the topological charge of the vortex beam $\ell$, and the propagation distance $z$ as in, 

\begin{equation}
    K = \dfrac{2^{3/2} \Gamma (\frac{1}{2} + |\ell|)}{\omega_z \pi |\ell|!}
\end{equation}

\noindent where $\omega_z$ is the beam radius at position $z$ along the optical axis~\cite{hermosa2011quadrant}. We observed that introducing the wobbling ratio $\beta$ produces asymmetric vortex beams, as shown in Figure \ref{fig:results} (a) and (b). Unlike the symmetric vortex beam, the slight displacement of the incident Gaussian beam results in a non-uniform intensity distribution which is more pronounced with large $\beta$ ($\beta \geq 10^{0}$). This asymmetry is hardly detectable with a CCD camera for small $\beta$ ($\beta << 10^0)$. Moreover, we found that the calibration response of an asymmetric vortex beams matches that of a uniform vortex beam for small values of $\beta$. For larger $\beta$ values, the calibration response transitions to resemble that of a Gaussian beam. We included the corresponding calibration value for each $\ell$ in our experimental measurements to account the response of the quadrant detector to these structured beams. 

\begin{figure}
	\centering
	\includegraphics[width=0.48\textwidth]{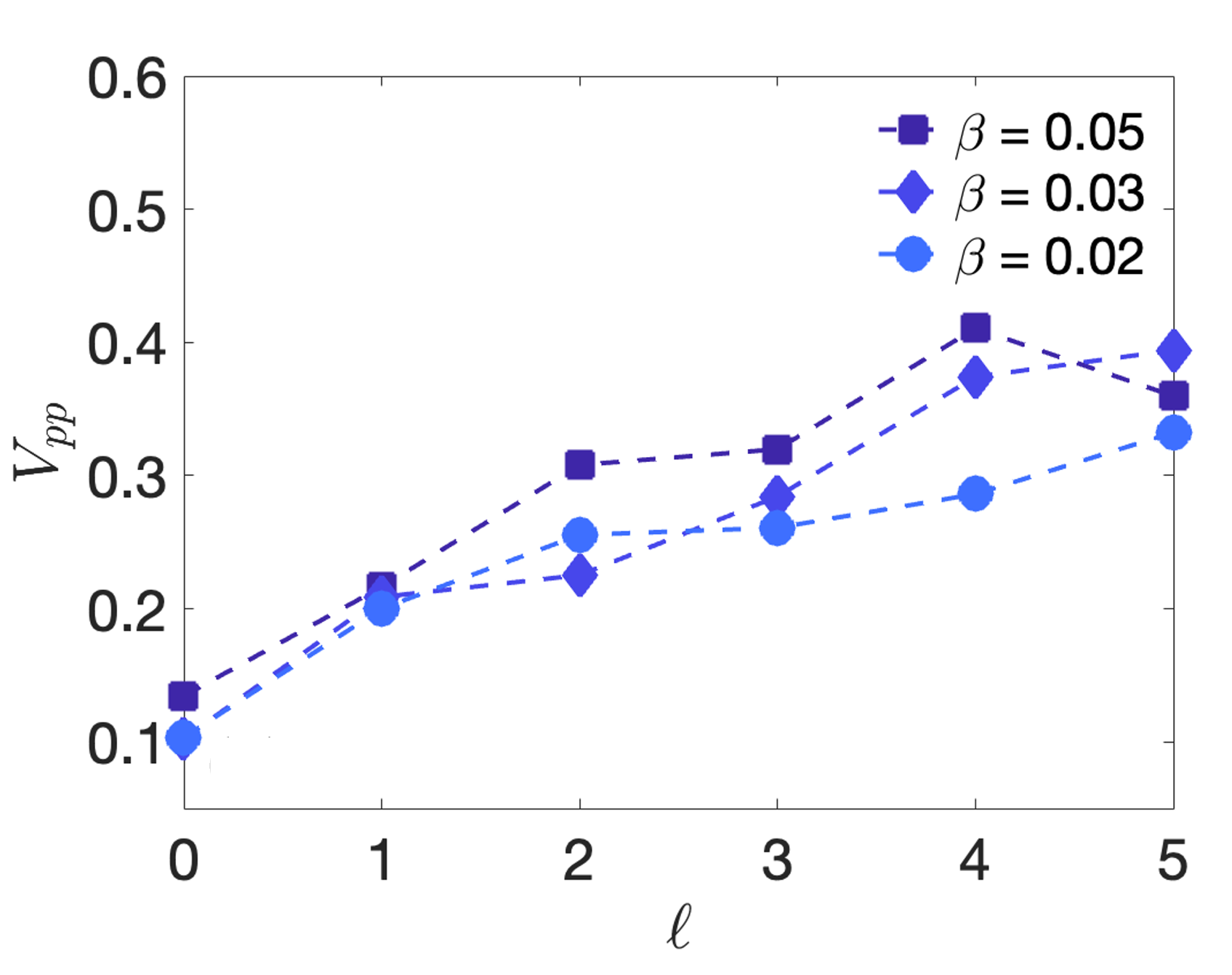}
	\caption{ We obtain the peak-to-peak value $V_{pp}$ of the normalized difference $\Delta V/V_{sum}$ from the voltage difference recorded in the oscilloscope. Using a vortex filter increases the peak-to-peak voltage signal of up to 400$\%$ compared to that of a regular Gaussian beam with $\ell = 0$. }
	\label{fig:Vpp}
\end{figure}

%results 
We configure the quadrant detector to function similarly to a split detector (i.e., $\Delta V = V_\text{right} - V_\text{left}$) and record the voltage difference using an oscilloscope.  The result is then normalized by the voltage sum $\Delta V/V_{sum}$, and we compute its peak-to-peak value $V_{pp}$. Figure \ref{fig:Vpp} presents the measured $V_{pp}$ values as a function of the topological charge $\ell$ for very small values of $\beta$ ($\beta <<1$) that we based on our analytical calculations. Comparing the result for $\ell = 5$ and a Gaussian beam $\ell = 0$, we observe an estimate increase of up to 400\% to the voltage signal. Our tabletop experiment serves a proof-of-concept; demonstrating that the use of a vortex filter enhances the signal from the quadrant detector.

%Since we have established that the calibration values for $\beta = 0$ and small $\beta$ values (e.g $\beta = 0.05$), we  use the  constant $K$ in equation \ref{eqn: constantK} to calibrate the recorded voltage signals . For small $\beta$ values, we observe an increase in the $V_{pp}$ signal when using higher $\ell$ values.  However, this is not the case for large $\beta$ values wherein a decrease was observed experimentally when we use higher order modes. 

%The analytical calculations agree well with numerical results as shown in Figure \ref{fig:ana_num}. As $\ell$ becomes larger, the peak-to-peak $\Delta I_\mathrm{norm}$ also increases for low $\beta$ values (Fig. \ref{fig:ana_num})$(a)$). The increase in $\Delta I_\mathrm{norm}$ is significant with a vortex beam compared to a Gaussian beam. For example, a phase plate with a charge $\ell=4$ gives rise to a  143.4\% increase in peak-to-peak signal. However, this is not the same trend for large $\beta$ where the $\Delta I_\mathrm{norm}$ do not have a considerable increase with $\ell$. From Figure \ref{fig:ana_num}(b) even for $\beta = 0.9$, the plots for different $\ell$ values are almost similar. From Equation \ref{eq:DeltaII}, it is obvious that there is a need to include more terms in the summation when $\beta$ is large. The number of terms to include depends on the value of $\beta$. In our calculations, we limit the $\beta$ values to be less than 1.

%% application
\section{Discussion}
 We show via analytical calculations and a proof-of-concept experiment that the $\Delta I_{\mathrm{norm}}$ from the wobble is enhanced by the $\ell$ of the vortex filter. The signal detected $\Delta I_{\mathrm{norm}}$ (Equation \ref{eq:DeltaII}) in small $\beta$ values is reduced to a simple sinusoidal function and the frequency $\omega$ can be determined. Based on our analytical calculations and experimental demonstration, our technique is specifically useful in detecting exoplanets whose wobbling ratio $\beta$ is in the order of $10^{-2}$ and lower. In our current experimental setup, the smallest achievable $\beta$ is $\beta = 0.02$ which is limited by the sensitivity of our quadrant detector. This corresponds to a displacement of $2\%$ of the incident beam's waist size. Analogously for an exoplanet in face-on orbit, the host star would be displaced by $2\%$ of its radius from the barycenter. We believe this limitation can be overcome by employing commercially available quadrant detectors with higher sensitivity.

This signal enhancement can still be increased by increasing the $\ell$ value. Unfortunately, however, there is a limit to this imposed by the size of the detector and the signal $\Delta I_{\mathrm{norm}}$ itself. The $\Delta I_{\mathrm{norm}}$ from Equation \ref{eq:DeltaII} will not increase indefinitely with $\ell$. At larger $\ell$ values, these equations tend to plateau and there will only be a marginal increase in the signal detected. Moreover, Padgett et al. in 2015 showed that vortex beam sizes scale $\propto \sqrt{\ell + 1 }$ for Gaussian beams impinging on diffractive optical elements \cite{padgett2015divergence}. A sufficiently large $\ell$ may form a vortex beam with dimensions greater than that of the detector area. 

\begin{figure}
	\centering
	\includegraphics[width=0.5\textwidth]{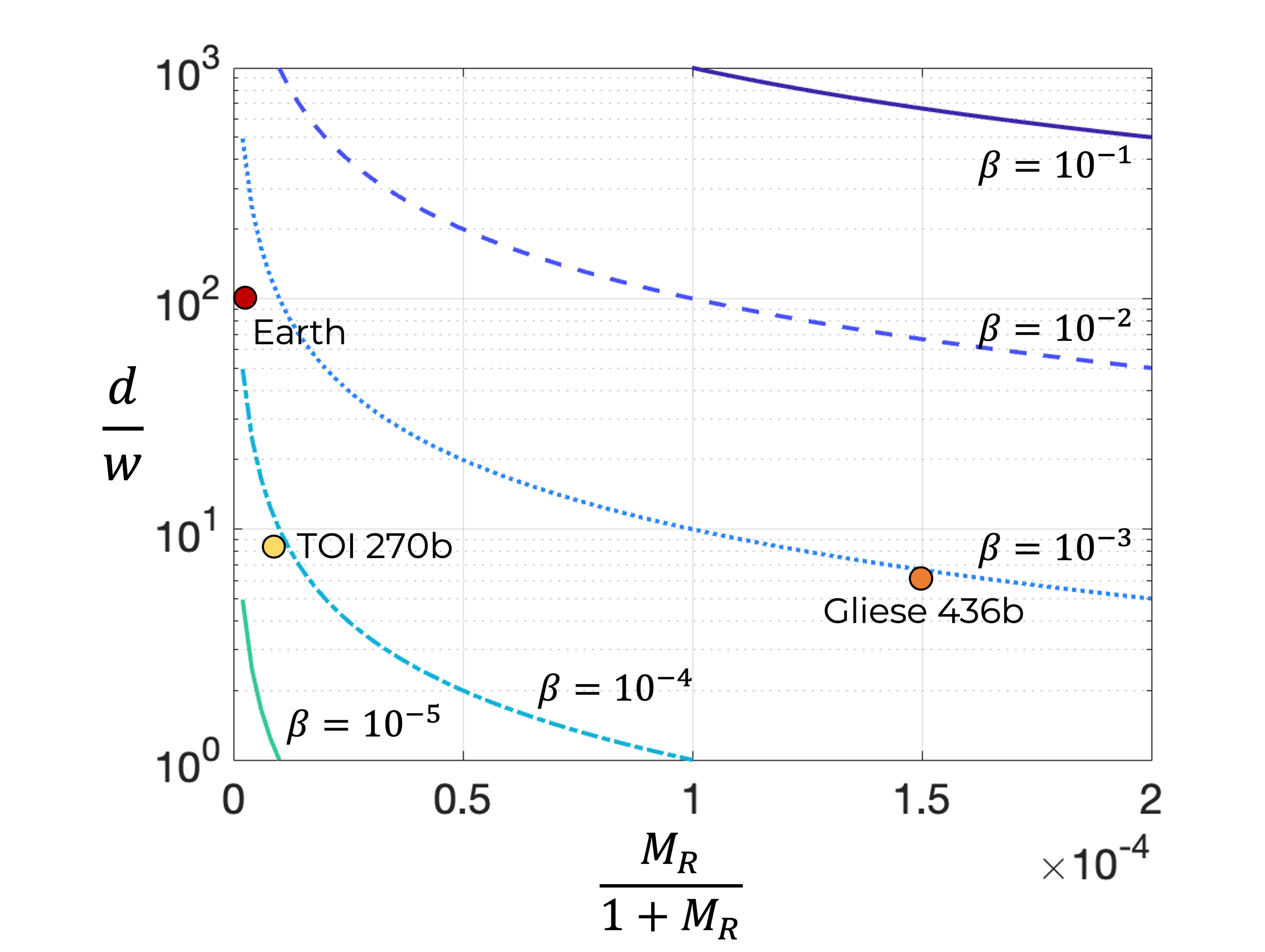}
	\caption{Wobbling ratio $\beta$ values for different $d/w$ and mass ratios $M_R$. Earth-like planets will have a $\beta \sim 10^{-3} - 10^{-4}$ with $d/w \sim 10^2$ and $M_R \sim 10^{-6}$. As a reference, our Earth is denoted by a red dot, and other Earth-like planets are also included.}
	\label{fig:wobble_mass}
\end{figure} 

We believe that our technique will be more successful in detecting exoplanets similar to Earth. Earth, together with other Earth-sized exoplanets, will have $\beta$ values in the $10^{-3}$ to $10^{-4}$ order, if Earth is the only planet in our solar system. Figure \ref{fig:wobble_mass} shows the $\beta$ values with respect to the ratio of the average distance of the planet $d$ to its host star with a diameter $w$, $d/w$ and the mass ratio of the planet to its host start, $M_R$ values. Exoplanets that are nearer and are much lighter to their host star are favored by our technique (Figure \ref{fig:wobble_mass}).

In observation, the $\beta$ value and hence, the exoplanet's mass $m_p$ and orbital distance $r_s$ can be calculated from $\Delta I_{\mathrm{norm}}$. As a simple calculation and assuming $\beta$ to be small, the peak-to-peak value is given by $2C(\beta,n=1,\ell)$. This value can then be used to detect the exoplanet's mass $m_p$ and its radius $r_p$ and its average distance from its host star $d$.  For our simplified case of a two-body system, we can exploit the periodicity of the $\Delta I_\mathrm{norm}$ plot to obtain frequency $\omega_s$ of the star around the barycenter.  The period $T_s$ taken by the star around the barycenter is  $T_s = (2\pi)/\omega_s$. Using Kepler's 3rd Law to relate the period $r_s=T_s^{2/3}\left(\frac{4\pi^2}{G m_s}\right)^{-1/3}$ where $T_s$ is the period and $r_s$ is the distance of the star around the barycenter and $G$ is the gravitational constant and $m_s$ is the mass of the host star, which can be known from other methods such as luminosity measurements. We set the orbital eccentricities to 0. The planet orbits the barycenter with the same period $T_p$ as that of the host star's $T_s$. Hence, we can get the $T_p$ that are functions of $r_p$ and $m_p$, $T_s=T_p=T= \left(\frac{4\pi^2}{G m_p}\right)^{1/2}r_p^{3/2}$. Since, $m_p$ and $r_p$ are still unknown variables, one can use the peak-to-peak intensity $2C(\beta,n=1,\ell)$, the $\beta$ value can be determined. For example for a fixed $\ell$ and at low $\beta$, $2C(\beta, n=1)  \propto \beta$ or is just a function of $\beta$. Once the $\beta$ value is obtained, it can be equated to $\beta = \frac{d}{w} \left(\frac{M_R}{1+M_R}\right)$ where $M_R=\frac{m_p}{m_s}$. With $m_p$ typically 3 to 4 orders of magnitude smaller than $m_s$, it can be simplified to be $\beta=\frac{d}{w_0} \left(\frac{M_R}{1+M_R}\right) \approx \frac{d}{w_0}\frac{m_p}{m_s}$ where $d$ is the average distance of the planet to its host star with a diameter $w_0$. With these equations, and knowing that $d=r_p +r_s$, we form a system of three equations with three unknowns: (1) $m_p = \left(\frac{4\pi^2}{GT^2}\right)r_p^3$; (2)
$d = (\beta w_0) \left(\frac{m_s}{m_p}\right)$; and (3) $d = T^{2/3}\left(\frac{4\pi^2}{G m_s}\right)^{-1/3} + r_p$. These three equations can be used to solve for the exoplanet mass $m_p$ and orbital distance $d$. Although the three equations presented here appear simple, solving for $r_p$, $m_p$, or $d$ will involve quartic equations that can be best solved numerically.

%At higher $\beta$ values, the signal is \emph{not} enhanced. Hence, we only calculated up to $\beta=0.9$. Since the planets with larger $\beta$ values will be relatively bigger and farther from their host star, it might be easier to detect them with coronagraphs or other imaging techniques. Our technique provides no advantage for large $\beta$.  Even exoplanets  that are Jupiter and Neptune-like with $\beta$ values of 0.518 and 0.165, respectively, may only marginally benefit from our technique. 

The lower limit of the detection entails discussion of the detector sensitivity. The main advantage of our technique is that it does not rely on images but only on the signal from the split detector. For a given order of magnitude of $\beta$, the $\Delta I_\mathrm{norm}$ peak-to-peak value has lower or at the same order of magnitude. For example, a $\beta$ value of $2\times10^{-4}$ to $5\times10^{-4}$ will have a $\Delta I_\mathrm{norm}$ peak-to-peak value of about $8\times10^{-4}$ to $4\times10^{-3}$, for $l=4$. Thus, to detect Super Earths or even Earth-sized exoplanets with $\beta$ values that are around this range, a photodetector capable of detecting intensity differences of 1 in a 10,000 will be needed. When we assume a pure Poissonian distribution of photons collected, the uncertainties related to the intensity at each panel of the split detector is $\sigma\approx\sqrt{I}$. This gives a rough SNR of $\frac{\Delta I_{norm}}{\sigma_1 + \sigma_2}\sim \frac{\sqrt{I_1}-\sqrt{I_2}}{I_{tot}}$, with each panel giving different uncertainty values.

In this study, we considered only the case where the barycenter of the two-body system is aligned with both the vortex filter and the split detector. To extend our calculations to an unknown barycenter, a general approach would be to position the vortex filter and split detector near the host star's center and adjust their alignment to reproduce signals similar to those in Figure \ref{fig:deltaI}.

\section{Conclusion}
In this work, we have demonstrated how vortex filters can enhance the astrometric detection of exoplanets with face-on orbits. We established our proposed technique through analytical calculations and a proof-of-concept experiment using a Gaussian beam incident on a spatial light modulator encoded with a vortex phase. Our results show that the peak-to-peak voltage signal, $V_{pp}$, increased by up to $400\%$ when the beam passed through an $\ell = 5$ phase filter compared to a Gaussian beam. This improvement highlights the potential of structured light for detecting exoplanets with small values of $\beta$ ($\beta \ll 1$), specifically planets that are relatively close to their host stars and have low mass ratios. In our current setup, the smallest achievable $\beta$ was $\beta = 0.02$, limited by the resolution of our quadrant detector. With advancements in space-based telescopes such as GAIA, our technique could offer a viable approach for improving astrometric precision in future astronomical observations.

\begin{acknowledgments}
We wish to acknowledge the support  of the DOST Philippine Council for Industry, Energy and Emerging Technology Research and Development (PCIEERD Project No. 04002),and the University of the Philippines Office of the Vice President for Academic Affairs thru its Balik-PhD Program (OVPAA-BPhD 2015-06) for financial support. NZ Simon would like to acknowledge the Office of the Chancellor of the University of the Philippines Diliman, through the Office of the Vice Chancellor for Research and Development, for funding support through the Thesis and Dissertation Grant (Project No. 232310 DNSE).  She is currently a postdoctoral fellow at the Toyota Technological Institute, Nagoya, Japan. M. Revilla is currently a PhD student at the Institute of Environmental Science and Meteorology, University of the Philippines Diliman, Diliman, Quezon City, Philippines, and a recipient of the DOST- Accelerated Science and Technology Human Resource Development Program (ASTHRDP) graduate scholarship . 
\end{acknowledgments}

% The \nocite command causes all entries in a bibliography to be printed out
% whether or not they are actually referenced in the text. This is appropriate
% for the sample file to show the different styles of references, but authors
% most likely will not want to use it.
%\nocite{*}

\bibliography{apssamp}% Produces the bibliography via BibTeX.

%apsrev4-2.bst 2019-01-14 (MD) hand-edited version of apsrev4-1.bst
%Control: key (0)
%Control: author (8) initials jnrlst
%Control: editor formatted (1) identically to author
%Control: production of article title (0) allowed
%Control: page (0) single
%Control: year (1) truncated
%Control: production of eprint (0) enabled
\begin{thebibliography}{30}%
\makeatletter
\providecommand \@ifxundefined [1]{%
 \@ifx{#1\undefined}
}%
\providecommand \@ifnum [1]{%
 \ifnum #1\expandafter \@firstoftwo
 \else \expandafter \@secondoftwo
 \fi
}%
\providecommand \@ifx [1]{%
 \ifx #1\expandafter \@firstoftwo
 \else \expandafter \@secondoftwo
 \fi
}%
\providecommand \natexlab [1]{#1}%
\providecommand \enquote  [1]{``#1''}%
\providecommand \bibnamefont  [1]{#1}%
\providecommand \bibfnamefont [1]{#1}%
\providecommand \citenamefont [1]{#1}%
\providecommand \href@noop [0]{\@secondoftwo}%
\providecommand \href [0]{\begingroup \@sanitize@url \@href}%
\providecommand \@href[1]{\@@startlink{#1}\@@href}%
\providecommand \@@href[1]{\endgroup#1\@@endlink}%
\providecommand \@sanitize@url [0]{\catcode `\\12\catcode `\$12\catcode
  `\&12\catcode `\#12\catcode `\^12\catcode `\_12\catcode `\%12\relax}%
\providecommand \@@startlink[1]{}%
\providecommand \@@endlink[0]{}%
\providecommand \url  [0]{\begingroup\@sanitize@url \@url }%
\providecommand \@url [1]{\endgroup\@href {#1}{\urlprefix }}%
\providecommand \urlprefix  [0]{URL }%
\providecommand \Eprint [0]{\href }%
\providecommand \doibase [0]{https://doi.org/}%
\providecommand \selectlanguage [0]{\@gobble}%
\providecommand \bibinfo  [0]{\@secondoftwo}%
\providecommand \bibfield  [0]{\@secondoftwo}%
\providecommand \translation [1]{[#1]}%
\providecommand \BibitemOpen [0]{}%
\providecommand \bibitemStop [0]{}%
\providecommand \bibitemNoStop [0]{.\EOS\space}%
\providecommand \EOS [0]{\spacefactor3000\relax}%
\providecommand \BibitemShut  [1]{\csname bibitem#1\endcsname}%
\let\auto@bib@innerbib\@empty
%</preamble>
\bibitem [{\citenamefont {Evans}(1998)}]{evans1998history}%
  \BibitemOpen
  \bibfield  {author} {\bibinfo {author} {\bibfnamefont {J.}~\bibnamefont
  {Evans}},\ }\href@noop {} {\emph {\bibinfo {title} {The history and practice
  of ancient astronomy}}}\ (\bibinfo  {publisher} {Oxford University Press},\
  \bibinfo {address} {Oxford},\ \bibinfo {year} {1998})\BibitemShut {NoStop}%
\bibitem [{\citenamefont {Sozzetti}\ and\ \citenamefont
  {de~Bruijne}(2018)}]{sozzetti2018space}%
  \BibitemOpen
  \bibfield  {author} {\bibinfo {author} {\bibfnamefont {A.}~\bibnamefont
  {Sozzetti}}\ and\ \bibinfo {author} {\bibfnamefont {J.}~\bibnamefont
  {de~Bruijne}},\ }\bibfield  {title} {\bibinfo {title} {Space astrometry
  missions for exoplanet science: Gaia and the legacy of hipparcos},\
  }\href@noop {} {\bibfield  {journal} {\bibinfo  {journal} {Handbook of
  Exoplanets}\ ,\ \bibinfo {pages} {81}} (\bibinfo {year} {2018})}\BibitemShut
  {NoStop}%
\bibitem [{\citenamefont {Launhardt}(2009)}]{launhardt2009exoplanet}%
  \BibitemOpen
  \bibfield  {author} {\bibinfo {author} {\bibfnamefont {R.}~\bibnamefont
  {Launhardt}},\ }\bibfield  {title} {\bibinfo {title} {Exoplanet search with
  astrometry},\ }\href@noop {} {\bibfield  {journal} {\bibinfo  {journal} {New
  Astronomy Reviews}\ }\textbf {\bibinfo {volume} {53}},\ \bibinfo {pages}
  {294} (\bibinfo {year} {2009})}\BibitemShut {NoStop}%
\bibitem [{\citenamefont {Sahlmann}\ \emph {et~al.}(2013)\citenamefont
  {Sahlmann}, \citenamefont {Lazorenko}, \citenamefont {S{\'e}gransan},
  \citenamefont {Mart{\'\i}n}, \citenamefont {Queloz}, \citenamefont {Mayor},\
  and\ \citenamefont {Udry}}]{sahlmann2013astrometric}%
  \BibitemOpen
  \bibfield  {author} {\bibinfo {author} {\bibfnamefont {J.}~\bibnamefont
  {Sahlmann}}, \bibinfo {author} {\bibfnamefont {P.}~\bibnamefont {Lazorenko}},
  \bibinfo {author} {\bibfnamefont {D.}~\bibnamefont {S{\'e}gransan}}, \bibinfo
  {author} {\bibfnamefont {E.}~\bibnamefont {Mart{\'\i}n}}, \bibinfo {author}
  {\bibfnamefont {D.}~\bibnamefont {Queloz}}, \bibinfo {author} {\bibfnamefont
  {M.}~\bibnamefont {Mayor}},\ and\ \bibinfo {author} {\bibfnamefont
  {S.}~\bibnamefont {Udry}},\ }\bibfield  {title} {\bibinfo {title}
  {Astrometric orbit of a low-mass companion to an ultracool dwarf},\
  }\href@noop {} {\bibfield  {journal} {\bibinfo  {journal} {Astronomy \&
  Astrophysics}\ }\textbf {\bibinfo {volume} {556}},\ \bibinfo {pages} {A133}
  (\bibinfo {year} {2013})}\BibitemShut {NoStop}%
\bibitem [{\citenamefont {Stef{\'a}nsson}\ \emph {et~al.}(2025)\citenamefont
  {Stef{\'a}nsson}, \citenamefont {Mahadevan}, \citenamefont {Winn},
  \citenamefont {Marcussen}, \citenamefont {Kanodia}, \citenamefont {Albrecht},
  \citenamefont {Fitzmaurice}, \citenamefont {Mikulskyt{\.e}}, \citenamefont
  {Ca{\~n}as}, \citenamefont {Espinoza-Retamal} \emph
  {et~al.}}]{stefansson2025gaia}%
  \BibitemOpen
  \bibfield  {author} {\bibinfo {author} {\bibfnamefont {G.}~\bibnamefont
  {Stef{\'a}nsson}}, \bibinfo {author} {\bibfnamefont {S.}~\bibnamefont
  {Mahadevan}}, \bibinfo {author} {\bibfnamefont {J.~N.}\ \bibnamefont {Winn}},
  \bibinfo {author} {\bibfnamefont {M.~L.}\ \bibnamefont {Marcussen}}, \bibinfo
  {author} {\bibfnamefont {S.}~\bibnamefont {Kanodia}}, \bibinfo {author}
  {\bibfnamefont {S.}~\bibnamefont {Albrecht}}, \bibinfo {author}
  {\bibfnamefont {E.}~\bibnamefont {Fitzmaurice}}, \bibinfo {author}
  {\bibfnamefont {O.}~\bibnamefont {Mikulskyt{\.e}}}, \bibinfo {author}
  {\bibfnamefont {C.~I.}\ \bibnamefont {Ca{\~n}as}}, \bibinfo {author}
  {\bibfnamefont {J.~I.}\ \bibnamefont {Espinoza-Retamal}}, \emph {et~al.},\
  }\bibfield  {title} {\bibinfo {title} {Gaia-4b and 5b: Radial velocity
  confirmation of gaia astrometric orbital solutions reveal a massive planet
  and a brown dwarf orbiting low-mass stars},\ }\href
  {https://doi.org/10.3847/1538-3881/ada9e1} {\bibfield  {journal} {\bibinfo
  {journal} {The Astronomical Journal}\ }\textbf {\bibinfo {volume} {169}},\
  \bibinfo {pages} {107} (\bibinfo {year} {2025})}\BibitemShut {NoStop}%
\bibitem [{\citenamefont {Malbet}\ and\ \citenamefont
  {Sozzetti}(2018)}]{malbet2018astrometry}%
  \BibitemOpen
  \bibfield  {author} {\bibinfo {author} {\bibfnamefont {F.}~\bibnamefont
  {Malbet}}\ and\ \bibinfo {author} {\bibfnamefont {A.}~\bibnamefont
  {Sozzetti}},\ }\bibfield  {title} {\bibinfo {title} {Astrometry as an
  exoplanet discovery method},\ }\href@noop {} {\bibfield  {journal} {\bibinfo
  {journal} {Handbook of Exoplanets}\ ,\ \bibinfo {pages} {196}} (\bibinfo
  {year} {2018})}\BibitemShut {NoStop}%
\bibitem [{\citenamefont {Quirrenbach}(2010)}]{quirrenbach2010astrometric}%
  \BibitemOpen
  \bibfield  {author} {\bibinfo {author} {\bibfnamefont {A.}~\bibnamefont
  {Quirrenbach}},\ }\bibfield  {title} {\bibinfo {title} {Astrometric detection
  and characterization of exoplanets},\ }\href@noop {} {\bibfield  {journal}
  {\bibinfo  {journal} {Exoplanets}\ ,\ \bibinfo {pages} {157}} (\bibinfo
  {year} {2010})}\BibitemShut {NoStop}%
\bibitem [{\citenamefont {Batalha}(2014)}]{batalha2014exploring}%
  \BibitemOpen
  \bibfield  {author} {\bibinfo {author} {\bibfnamefont {N.~M.}\ \bibnamefont
  {Batalha}},\ }\bibfield  {title} {\bibinfo {title} {Exploring exoplanet
  populations with nasa’s kepler mission},\ }\href@noop {} {\bibfield
  {journal} {\bibinfo  {journal} {Proceedings of the National Academy of
  Sciences}\ }\textbf {\bibinfo {volume} {111}},\ \bibinfo {pages} {12647}
  (\bibinfo {year} {2014})}\BibitemShut {NoStop}%
\bibitem [{\citenamefont {Deeg}\ and\ \citenamefont
  {Alonso}(2018)}]{deeg2018transit}%
  \BibitemOpen
  \bibfield  {author} {\bibinfo {author} {\bibfnamefont {H.~J.}\ \bibnamefont
  {Deeg}}\ and\ \bibinfo {author} {\bibfnamefont {R.}~\bibnamefont {Alonso}},\
  }\bibfield  {title} {\bibinfo {title} {Transit photometry as an exoplanet
  discovery method},\ }\href@noop {} {\bibfield  {journal} {\bibinfo  {journal}
  {Handbook of Exoplanets}\ ,\ \bibinfo {pages} {117}} (\bibinfo {year}
  {2018})}\BibitemShut {NoStop}%
\bibitem [{\citenamefont {Fischer}\ \emph {et~al.}(2014)\citenamefont
  {Fischer}, \citenamefont {Howard}, \citenamefont {Laughlin}, \citenamefont
  {Macintosh}, \citenamefont {Mahadevan}, \citenamefont {Sahlmann},\ and\
  \citenamefont {Yee}}]{fischer2014exoplanet}%
  \BibitemOpen
  \bibfield  {author} {\bibinfo {author} {\bibfnamefont {D.}~\bibnamefont
  {Fischer}}, \bibinfo {author} {\bibfnamefont {A.}~\bibnamefont {Howard}},
  \bibinfo {author} {\bibfnamefont {G.}~\bibnamefont {Laughlin}}, \bibinfo
  {author} {\bibfnamefont {B.}~\bibnamefont {Macintosh}}, \bibinfo {author}
  {\bibfnamefont {S.}~\bibnamefont {Mahadevan}}, \bibinfo {author}
  {\bibfnamefont {J.}~\bibnamefont {Sahlmann}},\ and\ \bibinfo {author}
  {\bibfnamefont {J.}~\bibnamefont {Yee}},\ }\bibfield  {title} {\bibinfo
  {title} {Exoplanet detection techniques},\ }\href@noop {} {\bibfield
  {journal} {\bibinfo  {journal} {Protostars and Planets VI}\ ,\ \bibinfo
  {pages} {715}} (\bibinfo {year} {2014})}\BibitemShut {NoStop}%
\bibitem [{\citenamefont {Hatzes}(2016)}]{hatzes2016radial}%
  \BibitemOpen
  \bibfield  {author} {\bibinfo {author} {\bibfnamefont {A.~P.}\ \bibnamefont
  {Hatzes}},\ }\bibfield  {title} {\bibinfo {title} {The radial velocity method
  for the detection of exoplanets},\ }in\ \href@noop {} {\emph {\bibinfo
  {booktitle} {Methods of Detecting Exoplanets}}}\ (\bibinfo  {publisher}
  {Springer},\ \bibinfo {address} {Cham},\ \bibinfo {year} {2016})\ pp.\
  \bibinfo {pages} {3--86}\BibitemShut {NoStop}%
\bibitem [{\citenamefont {Reiners}\ \emph {et~al.}(2010)\citenamefont
  {Reiners}, \citenamefont {Bean}, \citenamefont {Huber}, \citenamefont
  {Dreizler}, \citenamefont {Seifahrt},\ and\ \citenamefont
  {Czesla}}]{reiners2010detecting}%
  \BibitemOpen
  \bibfield  {author} {\bibinfo {author} {\bibfnamefont {A.}~\bibnamefont
  {Reiners}}, \bibinfo {author} {\bibfnamefont {J.}~\bibnamefont {Bean}},
  \bibinfo {author} {\bibfnamefont {K.}~\bibnamefont {Huber}}, \bibinfo
  {author} {\bibfnamefont {S.}~\bibnamefont {Dreizler}}, \bibinfo {author}
  {\bibfnamefont {A.}~\bibnamefont {Seifahrt}},\ and\ \bibinfo {author}
  {\bibfnamefont {S.}~\bibnamefont {Czesla}},\ }\bibfield  {title} {\bibinfo
  {title} {Detecting planets around very low mass stars with the radial
  velocity method},\ }\href@noop {} {\bibfield  {journal} {\bibinfo  {journal}
  {The Astrophysical Journal}\ }\textbf {\bibinfo {volume} {710}},\ \bibinfo
  {pages} {432} (\bibinfo {year} {2010})}\BibitemShut {NoStop}%
\bibitem [{\citenamefont {Knutson}\ \emph {et~al.}(2014)\citenamefont
  {Knutson}, \citenamefont {Fulton}, \citenamefont {Montet}, \citenamefont
  {Kao}, \citenamefont {Ngo}, \citenamefont {Howard}, \citenamefont {Crepp},
  \citenamefont {Hinkley}, \citenamefont {Bakos}, \citenamefont {Batygin} \emph
  {et~al.}}]{knutson2014friends}%
  \BibitemOpen
  \bibfield  {author} {\bibinfo {author} {\bibfnamefont {H.~A.}\ \bibnamefont
  {Knutson}}, \bibinfo {author} {\bibfnamefont {B.~J.}\ \bibnamefont {Fulton}},
  \bibinfo {author} {\bibfnamefont {B.~T.}\ \bibnamefont {Montet}}, \bibinfo
  {author} {\bibfnamefont {M.}~\bibnamefont {Kao}}, \bibinfo {author}
  {\bibfnamefont {H.}~\bibnamefont {Ngo}}, \bibinfo {author} {\bibfnamefont
  {A.~W.}\ \bibnamefont {Howard}}, \bibinfo {author} {\bibfnamefont {J.~R.}\
  \bibnamefont {Crepp}}, \bibinfo {author} {\bibfnamefont {S.}~\bibnamefont
  {Hinkley}}, \bibinfo {author} {\bibfnamefont {G.~{\'A}.}\ \bibnamefont
  {Bakos}}, \bibinfo {author} {\bibfnamefont {K.}~\bibnamefont {Batygin}},
  \emph {et~al.},\ }\bibfield  {title} {\bibinfo {title} {Friends of hot
  jupiters. i. a radial velocity search for massive, long-period companions to
  close-in gas giant planets},\ }\href@noop {} {\bibfield  {journal} {\bibinfo
  {journal} {The Astrophysical Journal}\ }\textbf {\bibinfo {volume} {785}},\
  \bibinfo {pages} {126} (\bibinfo {year} {2014})}\BibitemShut {NoStop}%
\bibitem [{\citenamefont {Hecht}(2002)}]{hecht2002optics}%
  \BibitemOpen
  \bibfield  {author} {\bibinfo {author} {\bibfnamefont {E.}~\bibnamefont
  {Hecht}},\ }\href@noop {} {\emph {\bibinfo {title} {Optics, 5th Ed}}}\
  (\bibinfo  {publisher} {Pearson Education, Inc.},\ \bibinfo {address} {San
  Francisco},\ \bibinfo {year} {2002})\BibitemShut {NoStop}%
\bibitem [{\citenamefont {Pueyo}(2018)}]{pueyo2018direct}%
  \BibitemOpen
  \bibfield  {author} {\bibinfo {author} {\bibfnamefont {L.}~\bibnamefont
  {Pueyo}},\ }\bibfield  {title} {\bibinfo {title} {Direct imaging as a
  detection technique for exoplanets},\ }\href@noop {} {\bibfield  {journal}
  {\bibinfo  {journal} {Handbook of Exoplanets}\ ,\ \bibinfo {pages} {10}}
  (\bibinfo {year} {2018})}\BibitemShut {NoStop}%
\bibitem [{\citenamefont {Foo}\ \emph {et~al.}(2005)\citenamefont {Foo},
  \citenamefont {Palacios},\ and\ \citenamefont
  {Swartzlander}}]{foo2005optical}%
  \BibitemOpen
  \bibfield  {author} {\bibinfo {author} {\bibfnamefont {G.}~\bibnamefont
  {Foo}}, \bibinfo {author} {\bibfnamefont {D.~M.}\ \bibnamefont {Palacios}},\
  and\ \bibinfo {author} {\bibfnamefont {G.~A.}\ \bibnamefont {Swartzlander}},\
  }\bibfield  {title} {\bibinfo {title} {Optical vortex coronagraph},\
  }\href@noop {} {\bibfield  {journal} {\bibinfo  {journal} {Optics letters}\
  }\textbf {\bibinfo {volume} {30}},\ \bibinfo {pages} {3308} (\bibinfo {year}
  {2005})}\BibitemShut {NoStop}%
\bibitem [{\citenamefont {Tamburini}\ \emph {et~al.}(2006)\citenamefont
  {Tamburini}, \citenamefont {Anzolin}, \citenamefont {Umbriaco}, \citenamefont
  {Bianchini},\ and\ \citenamefont {Barbieri}}]{tamburini2006overcoming}%
  \BibitemOpen
  \bibfield  {author} {\bibinfo {author} {\bibfnamefont {F.}~\bibnamefont
  {Tamburini}}, \bibinfo {author} {\bibfnamefont {G.}~\bibnamefont {Anzolin}},
  \bibinfo {author} {\bibfnamefont {G.}~\bibnamefont {Umbriaco}}, \bibinfo
  {author} {\bibfnamefont {A.}~\bibnamefont {Bianchini}},\ and\ \bibinfo
  {author} {\bibfnamefont {C.}~\bibnamefont {Barbieri}},\ }\bibfield  {title}
  {\bibinfo {title} {Overcoming the rayleigh criterion limit with optical
  vortices},\ }\href@noop {} {\bibfield  {journal} {\bibinfo  {journal}
  {Physical review letters}\ }\textbf {\bibinfo {volume} {97}},\ \bibinfo
  {pages} {163903} (\bibinfo {year} {2006})}\BibitemShut {NoStop}%
\bibitem [{\citenamefont {Mawet}\ \emph {et~al.}(2005)\citenamefont {Mawet},
  \citenamefont {Riaud}, \citenamefont {Absil},\ and\ \citenamefont
  {Surdej}}]{mawet2005annular}%
  \BibitemOpen
  \bibfield  {author} {\bibinfo {author} {\bibfnamefont {D.}~\bibnamefont
  {Mawet}}, \bibinfo {author} {\bibfnamefont {P.}~\bibnamefont {Riaud}},
  \bibinfo {author} {\bibfnamefont {O.}~\bibnamefont {Absil}},\ and\ \bibinfo
  {author} {\bibfnamefont {J.}~\bibnamefont {Surdej}},\ }\bibfield  {title}
  {\bibinfo {title} {Annular groove phase mask coronagraph},\ }\href@noop {}
  {\bibfield  {journal} {\bibinfo  {journal} {The Astrophysical Journal}\
  }\textbf {\bibinfo {volume} {633}},\ \bibinfo {pages} {1191} (\bibinfo {year}
  {2005})}\BibitemShut {NoStop}%
\bibitem [{\citenamefont {Serabyn}\ \emph {et~al.}(2010)\citenamefont
  {Serabyn}, \citenamefont {Mawet},\ and\ \citenamefont
  {Burruss}}]{serabyn2010image}%
  \BibitemOpen
  \bibfield  {author} {\bibinfo {author} {\bibfnamefont {E.}~\bibnamefont
  {Serabyn}}, \bibinfo {author} {\bibfnamefont {D.}~\bibnamefont {Mawet}},\
  and\ \bibinfo {author} {\bibfnamefont {R.}~\bibnamefont {Burruss}},\
  }\bibfield  {title} {\bibinfo {title} {An image of an exoplanet separated by
  two diffraction beamwidths from a star},\ }\href@noop {} {\bibfield
  {journal} {\bibinfo  {journal} {Nature}\ }\textbf {\bibinfo {volume} {464}},\
  \bibinfo {pages} {1018} (\bibinfo {year} {2010})}\BibitemShut {NoStop}%
\bibitem [{\citenamefont {Wagner}\ \emph {et~al.}(2021)\citenamefont {Wagner},
  \citenamefont {Boehle}, \citenamefont {Pathak}, \citenamefont {Kasper},
  \citenamefont {Arsenault}, \citenamefont {Jakob}, \citenamefont {K{\"a}ufl},
  \citenamefont {Leveratto}, \citenamefont {Maire}, \citenamefont {Pantin}
  \emph {et~al.}}]{wagner2021imaging}%
  \BibitemOpen
  \bibfield  {author} {\bibinfo {author} {\bibfnamefont {K.}~\bibnamefont
  {Wagner}}, \bibinfo {author} {\bibfnamefont {A.}~\bibnamefont {Boehle}},
  \bibinfo {author} {\bibfnamefont {P.}~\bibnamefont {Pathak}}, \bibinfo
  {author} {\bibfnamefont {M.}~\bibnamefont {Kasper}}, \bibinfo {author}
  {\bibfnamefont {R.}~\bibnamefont {Arsenault}}, \bibinfo {author}
  {\bibfnamefont {G.}~\bibnamefont {Jakob}}, \bibinfo {author} {\bibfnamefont
  {U.}~\bibnamefont {K{\"a}ufl}}, \bibinfo {author} {\bibfnamefont
  {S.}~\bibnamefont {Leveratto}}, \bibinfo {author} {\bibfnamefont {A.-L.}\
  \bibnamefont {Maire}}, \bibinfo {author} {\bibfnamefont {E.}~\bibnamefont
  {Pantin}}, \emph {et~al.},\ }\bibfield  {title} {\bibinfo {title} {Imaging
  low-mass planets within the habitable zone of $\alpha$ centauri},\
  }\href@noop {} {\bibfield  {journal} {\bibinfo  {journal} {Nature
  communications}\ }\textbf {\bibinfo {volume} {12}},\ \bibinfo {pages} {1}
  (\bibinfo {year} {2021})}\BibitemShut {NoStop}%
\bibitem [{\citenamefont {Guyon}\ \emph {et~al.}(2006)\citenamefont {Guyon},
  \citenamefont {Pluzhnik}, \citenamefont {Kuchner}, \citenamefont {Collins},\
  and\ \citenamefont {Ridgway}}]{guyon2006theoretical}%
  \BibitemOpen
  \bibfield  {author} {\bibinfo {author} {\bibfnamefont {O.}~\bibnamefont
  {Guyon}}, \bibinfo {author} {\bibfnamefont {E.}~\bibnamefont {Pluzhnik}},
  \bibinfo {author} {\bibfnamefont {M.~J.}\ \bibnamefont {Kuchner}}, \bibinfo
  {author} {\bibfnamefont {B.}~\bibnamefont {Collins}},\ and\ \bibinfo {author}
  {\bibfnamefont {S.}~\bibnamefont {Ridgway}},\ }\bibfield  {title} {\bibinfo
  {title} {Theoretical limits on extrasolar terrestrial planet detection with
  coronagraphs},\ }\href@noop {} {\bibfield  {journal} {\bibinfo  {journal}
  {The Astrophysical Journal Supplement Series}\ }\textbf {\bibinfo {volume}
  {167}},\ \bibinfo {pages} {81} (\bibinfo {year} {2006})}\BibitemShut
  {NoStop}%
\bibitem [{\citenamefont {Mawet}\ \emph {et~al.}(2014)\citenamefont {Mawet},
  \citenamefont {Milli}, \citenamefont {Wahhaj}, \citenamefont {Pelat},
  \citenamefont {Absil}, \citenamefont {Delacroix}, \citenamefont {Boccaletti},
  \citenamefont {Kasper}, \citenamefont {Kenworthy}, \citenamefont {Marois}
  \emph {et~al.}}]{mawet2014fundamental}%
  \BibitemOpen
  \bibfield  {author} {\bibinfo {author} {\bibfnamefont {D.}~\bibnamefont
  {Mawet}}, \bibinfo {author} {\bibfnamefont {J.}~\bibnamefont {Milli}},
  \bibinfo {author} {\bibfnamefont {Z.}~\bibnamefont {Wahhaj}}, \bibinfo
  {author} {\bibfnamefont {D.}~\bibnamefont {Pelat}}, \bibinfo {author}
  {\bibfnamefont {O.}~\bibnamefont {Absil}}, \bibinfo {author} {\bibfnamefont
  {C.}~\bibnamefont {Delacroix}}, \bibinfo {author} {\bibfnamefont
  {A.}~\bibnamefont {Boccaletti}}, \bibinfo {author} {\bibfnamefont
  {M.}~\bibnamefont {Kasper}}, \bibinfo {author} {\bibfnamefont
  {M.}~\bibnamefont {Kenworthy}}, \bibinfo {author} {\bibfnamefont
  {C.}~\bibnamefont {Marois}}, \emph {et~al.},\ }\bibfield  {title} {\bibinfo
  {title} {Fundamental limitations of high contrast imaging set by small sample
  statistics},\ }\href@noop {} {\bibfield  {journal} {\bibinfo  {journal} {The
  Astrophysical Journal}\ }\textbf {\bibinfo {volume} {792}},\ \bibinfo {pages}
  {97} (\bibinfo {year} {2014})}\BibitemShut {NoStop}%
\bibitem [{\citenamefont {Hermosa}\ \emph {et~al.}(2011)\citenamefont
  {Hermosa}, \citenamefont {Aiello},\ and\ \citenamefont
  {Woerdman}}]{hermosa2011quadrant}%
  \BibitemOpen
  \bibfield  {author} {\bibinfo {author} {\bibfnamefont {N.}~\bibnamefont
  {Hermosa}}, \bibinfo {author} {\bibfnamefont {A.}~\bibnamefont {Aiello}},\
  and\ \bibinfo {author} {\bibfnamefont {J.}~\bibnamefont {Woerdman}},\
  }\bibfield  {title} {\bibinfo {title} {Quadrant detector calibration for
  vortex beams},\ }\href@noop {} {\bibfield  {journal} {\bibinfo  {journal}
  {Optics letters}\ }\textbf {\bibinfo {volume} {36}},\ \bibinfo {pages} {409}
  (\bibinfo {year} {2011})}\BibitemShut {NoStop}%
\bibitem [{\citenamefont {Narag}\ and\ \citenamefont
  {Hermosa}(2017)}]{narag2017response}%
  \BibitemOpen
  \bibfield  {author} {\bibinfo {author} {\bibfnamefont {J.}~\bibnamefont
  {Narag}}\ and\ \bibinfo {author} {\bibfnamefont {N.}~\bibnamefont
  {Hermosa}},\ }\bibfield  {title} {\bibinfo {title} {Response of quadrant
  detectors to structured beams via convolution integrals},\ }\href@noop {}
  {\bibfield  {journal} {\bibinfo  {journal} {JOSA A}\ }\textbf {\bibinfo
  {volume} {34}},\ \bibinfo {pages} {1212} (\bibinfo {year}
  {2017})}\BibitemShut {NoStop}%
\bibitem [{\citenamefont {Thid{\'e}}\ \emph {et~al.}(2011)\citenamefont
  {Thid{\'e}}, \citenamefont {Elias}, \citenamefont {Tamburini}, \citenamefont
  {Mohammadi},\ and\ \citenamefont {Mendon{\c{c}}a}}]{thide2011applications}%
  \BibitemOpen
  \bibfield  {author} {\bibinfo {author} {\bibfnamefont {B.}~\bibnamefont
  {Thid{\'e}}}, \bibinfo {author} {\bibfnamefont {N.~M.}\ \bibnamefont
  {Elias}}, \bibinfo {author} {\bibfnamefont {F.}~\bibnamefont {Tamburini}},
  \bibinfo {author} {\bibfnamefont {S.~M.}\ \bibnamefont {Mohammadi}},\ and\
  \bibinfo {author} {\bibfnamefont {J.~T.}\ \bibnamefont {Mendon{\c{c}}a}},\
  }\bibfield  {title} {\bibinfo {title} {Applications of electromagnetic oam in
  astrophysics and space physics studies},\ }\href@noop {} {\bibfield
  {journal} {\bibinfo  {journal} {Twisted photons: applications of light with
  orbital angular momentum}\ ,\ \bibinfo {pages} {155}} (\bibinfo {year}
  {2011})}\BibitemShut {NoStop}%
\bibitem [{\citenamefont {Anzolin}\ \emph {et~al.}(2009)\citenamefont
  {Anzolin}, \citenamefont {Tamburini}, \citenamefont {Bianchini},\ and\
  \citenamefont {Barbieri}}]{anzolin2009method}%
  \BibitemOpen
  \bibfield  {author} {\bibinfo {author} {\bibfnamefont {G.}~\bibnamefont
  {Anzolin}}, \bibinfo {author} {\bibfnamefont {F.}~\bibnamefont {Tamburini}},
  \bibinfo {author} {\bibfnamefont {A.}~\bibnamefont {Bianchini}},\ and\
  \bibinfo {author} {\bibfnamefont {C.}~\bibnamefont {Barbieri}},\ }\bibfield
  {title} {\bibinfo {title} {Method to measure off-axis displacements based on
  the analysis of the intensity distribution of a vortex beam},\ }\href@noop {}
  {\bibfield  {journal} {\bibinfo  {journal} {Physical Review A}\ }\textbf
  {\bibinfo {volume} {79}},\ \bibinfo {pages} {033845} (\bibinfo {year}
  {2009})}\BibitemShut {NoStop}%
\bibitem [{\citenamefont {Kotlyar}\ \emph {et~al.}(2017)\citenamefont
  {Kotlyar}, \citenamefont {Kovalev},\ and\ \citenamefont
  {Porfirev}}]{asymmetricgaussian}%
  \BibitemOpen
  \bibfield  {author} {\bibinfo {author} {\bibfnamefont {V.}~\bibnamefont
  {Kotlyar}}, \bibinfo {author} {\bibfnamefont {A.}~\bibnamefont {Kovalev}},\
  and\ \bibinfo {author} {\bibfnamefont {A.}~\bibnamefont {Porfirev}},\
  }\bibfield  {title} {\bibinfo {title} {Asymmetric gaussian optical vortex},\
  }\href@noop {} {\bibfield  {journal} {\bibinfo  {journal} {Opt. Lett.}\
  }\textbf {\bibinfo {volume} {42}},\ \bibinfo {pages} {139} (\bibinfo {year}
  {2017})}\BibitemShut {NoStop}%
\bibitem [{\citenamefont {Bekshaev}\ and\ \citenamefont
  {Karamoch}(2008)}]{bekshaev2008spatial}%
  \BibitemOpen
  \bibfield  {author} {\bibinfo {author} {\bibfnamefont {A.~Y.}\ \bibnamefont
  {Bekshaev}}\ and\ \bibinfo {author} {\bibfnamefont {A.}~\bibnamefont
  {Karamoch}},\ }\bibfield  {title} {\bibinfo {title} {Spatial characteristics
  of vortex light beams produced by diffraction gratings with embedded phase
  singularity},\ }\href@noop {} {\bibfield  {journal} {\bibinfo  {journal}
  {Optics Communications}\ }\textbf {\bibinfo {volume} {281}},\ \bibinfo
  {pages} {1366} (\bibinfo {year} {2008})}\BibitemShut {NoStop}%
\bibitem [{\citenamefont {Abramowitz}\ and\ \citenamefont
  {Stegun}(1972)}]{AbraSteg}%
  \BibitemOpen
  \bibfield  {author} {\bibinfo {author} {\bibfnamefont {M.}~\bibnamefont
  {Abramowitz}}\ and\ \bibinfo {author} {\bibfnamefont {I.}~\bibnamefont
  {Stegun}},\ }\href@noop {} {\emph {\bibinfo {title} {Handbook of
  {M}athematical {F}unctions with {F}ormulas, {G}raphs, and {M}athematical
  {T}ables}}}\ (\bibinfo  {publisher} {National Bureau of Standards},\ \bibinfo
  {year} {1972})\BibitemShut {NoStop}%
\bibitem [{\citenamefont {Padgett}\ \emph {et~al.}(2015)\citenamefont
  {Padgett}, \citenamefont {Miatto}, \citenamefont {Lavery}, \citenamefont
  {Zeilinger},\ and\ \citenamefont {Boyd}}]{padgett2015divergence}%
  \BibitemOpen
  \bibfield  {author} {\bibinfo {author} {\bibfnamefont {M.~J.}\ \bibnamefont
  {Padgett}}, \bibinfo {author} {\bibfnamefont {F.~M.}\ \bibnamefont {Miatto}},
  \bibinfo {author} {\bibfnamefont {M.~P.}\ \bibnamefont {Lavery}}, \bibinfo
  {author} {\bibfnamefont {A.}~\bibnamefont {Zeilinger}},\ and\ \bibinfo
  {author} {\bibfnamefont {R.~W.}\ \bibnamefont {Boyd}},\ }\bibfield  {title}
  {\bibinfo {title} {Divergence of an orbital-angular-momentum-carrying beam
  upon propagation},\ }\href@noop {} {\bibfield  {journal} {\bibinfo  {journal}
  {New Journal of Physics}\ }\textbf {\bibinfo {volume} {17}},\ \bibinfo
  {pages} {023011} (\bibinfo {year} {2015})}\BibitemShut {NoStop}%
\end{thebibliography}%

\end{document}